\documentclass[prb,twocolumn]{revtex4}%

\usepackage{graphicx}
\usepackage{amsmath}
\usepackage{ulem}
\usepackage{color}

\newcommand{\be}{\begin{equation}}
\newcommand{\ee}{\end{equation}}
\newcommand{\bea}{\begin{eqnarray*}}
\newcommand{\eea}{\end{eqnarray*}}
\newcommand{\bean}{\begin{eqnarray}}
\newcommand{\eean}{\end{eqnarray}}

\begin{document}

\draft
\title{\bf Effects of zigzag edge states on the
thermoelectric properties of finite graphene nanoribbons}

\author{David Ming Ting Kuo
}

\address{Department of Electrical Engineering and Department of Physics, National Central
University, Chungli, 320 Taiwan}

%\address{$^{2}$Research Center for Applied Sciences, Academic Sinica,
%Taipei, 11529 Taiwan}

\date{\today}

\begin{abstract}
Thermoelectric properties of finite graphene nanoribbons (GNRs)
coupled to metallic electrodes are theoretically studied in the
framework of tight-binding model and Green's function approach.
When the zigzag sides are coupled to the electrodes, the electron
transport through the localized edge states can occur only if the
channel length between electrodes is smaller than the decay length
of these localized zigzag edge states. When the armchair edges are
coupled to the electrodes, there is an interesting thermoelectric
behavior associated with the mid-gap states when the GNR is in the
semiconducting phase. Here we show that the thermoelectric
behavior of zigzag edge states of GNRs with armchair sides
connected to electrodes is similar to that of two parallel quantum
dots with similar orbital degeneracy. Furthermore, it is
demonstrated that the electrical conductance and power factor
given by the zigzag edge states are quite robust against the
defect scattering.
\end{abstract}

\maketitle

\section{Introduction}

The efficiency of thermoelectric materials is determined by the
dimensionless figure of merit $ZT=S^2G_eT/(\kappa_e+\kappa_{ph})$,
which depends on the Seeback coefficient ($S$), electrical
conductance ($G_e$) and thermal conductance
($\kappa=\kappa_e+\kappa_{ph}$) of the material.$^{1,2)}$ The
thermal conductance includes contributions due to electron
transport ($\kappa_e$) and phonon transport ($\kappa_{ph}$). Here,
$T$ denotes the equilibrium temperature of the thermoelectric
device. For applications of thermoelectric devices, not only the
thermoelectric efficiency but also the electrical power output
need to be optimized. However, there is often a trade off between
the efficiency and power output in conventional thermoelectric
materials.$^{2)}$ For example, with the increase of $G_e$ one
often faces the increase of $\kappa_e$ and reduction of $S$. As a
result $ZT$ is reduced.

To find the best compromise between thermoelectric efficiency and
electrical power output, Hicks and Dresselhaus theoretically
demonstrated that the thermoelectric performance  can be
significantly enhanced in low-dimensional systems due to the
reduced $\kappa_{ph}$.$^{3)}$ For quasi one dimensional systems,
$\kappa_{ph}$ can be highly reduced meanwhile the power factor,
$PF=S^2G_e$ could remain similar to that of bulk materials due to
the enhancement of $S$.$^{4)}$ Nanowires with high thermoelectric
efficiency have been reported in several theoretical and
experimental studies.$^{4-8)}$ Nevertheless, these quasi
one-dimensional nanowires have about a few hundred atoms in the
cross-sectional area with diameter of $10~nm$ for the carrier
transport.$^{4-8)}$ The fabrication of one-dimensional (1D) solid
state system with smaller cross-section was a challenging problem
in material science. The discovery of two-dimensional (2D)
graphene in 2004 opened the door for realizing 1D systems with
small cross-section,$^{9)}$ since one can atomically precise
fabricate graphene nanoribbons (GNRs) by bottom-up
approach.$^{10)}$ Graphene with a hexagonal lattice structure is a
zero-gap semiconductor because its conduction and valence bands
meet at the Dirac points.$^{11)}$ Although such a zero-gap
property limits its applications in electronics and
optoelectronics, the discovery of 2D graphene stimulates an
impressive development of 2D materials with semiconducting
properties.

In recent years, tremendous efforts have been devoted to the
investigation of the transport and optical properties of 2D
materials for applications of new generation electronics and
optoelectronics.$^{12-21)}$ For nanowires realized by 2D
materials, it is expected that their thermal conductances will be
dramatically reduced owning to the small number of atoms in the
cross section.$^{22-25)}$ Compared with nanoribbons made from
other two-dimensional material, GNRs are easier to
obtain.$^{10,26,27)}$Meanwhile, recycling carbons plays an
important role to eternal development of the earth. Therefore, it
is desirable to clarify the transport and thermoelectric
properties of GNRs.$^{22,28,29)}$ GNRs have zigzag GRNs (ZGRNs)
and armchair GRNs (AGNRs). Because AGNRs exhibit semiconductor
phases, most theoretical studies have focused on thermoelectric
performance of AGRNs.$^{22,28,29)}$ When the zigzag sides of an
infinitely long AGRNs are coupled to the electrodes, electron
transport through the zigzag-edge states is suppressed due to
their localize wave functions.$^{30-36)}$ For finite-size GNRs,
there exists quantum-confinement effect which depends on whether
the short sides of the GNR have zigzag or armchair edges. For
experimentally studied AGNRs, the edge states are always present
and such a study could provide more insight into the transport
characteristics. Size effects on electronic structures of
finite-size GNRs have been studied theoretically,$^{37,38)}$
however the investigation of their thermoelectrical properties is
still lacking.

Here we theoretically investigate the transport and thermoelectric
properties of finite size GNRs coupled to the metallic electrodes
with two kinds of geometries as shown in Fig. 1. Electron
transport is determined not only by GRN types but also their
contact geometries. The finite size GNRs in the configuration as
shown in Fig. 1(b) could provide a sharp transmission coefficient
locating at middle of band gap, which is created by zigzag edges
of GNRs. The main goal of this study is to illustrate how zigzag
edge states to influence the thermoelectric properties of GNRs.

\begin{figure}[h]
\centering
\includegraphics[trim=2.5cm 0cm 2.5cm 0cm,clip,angle=0,scale=0.3]{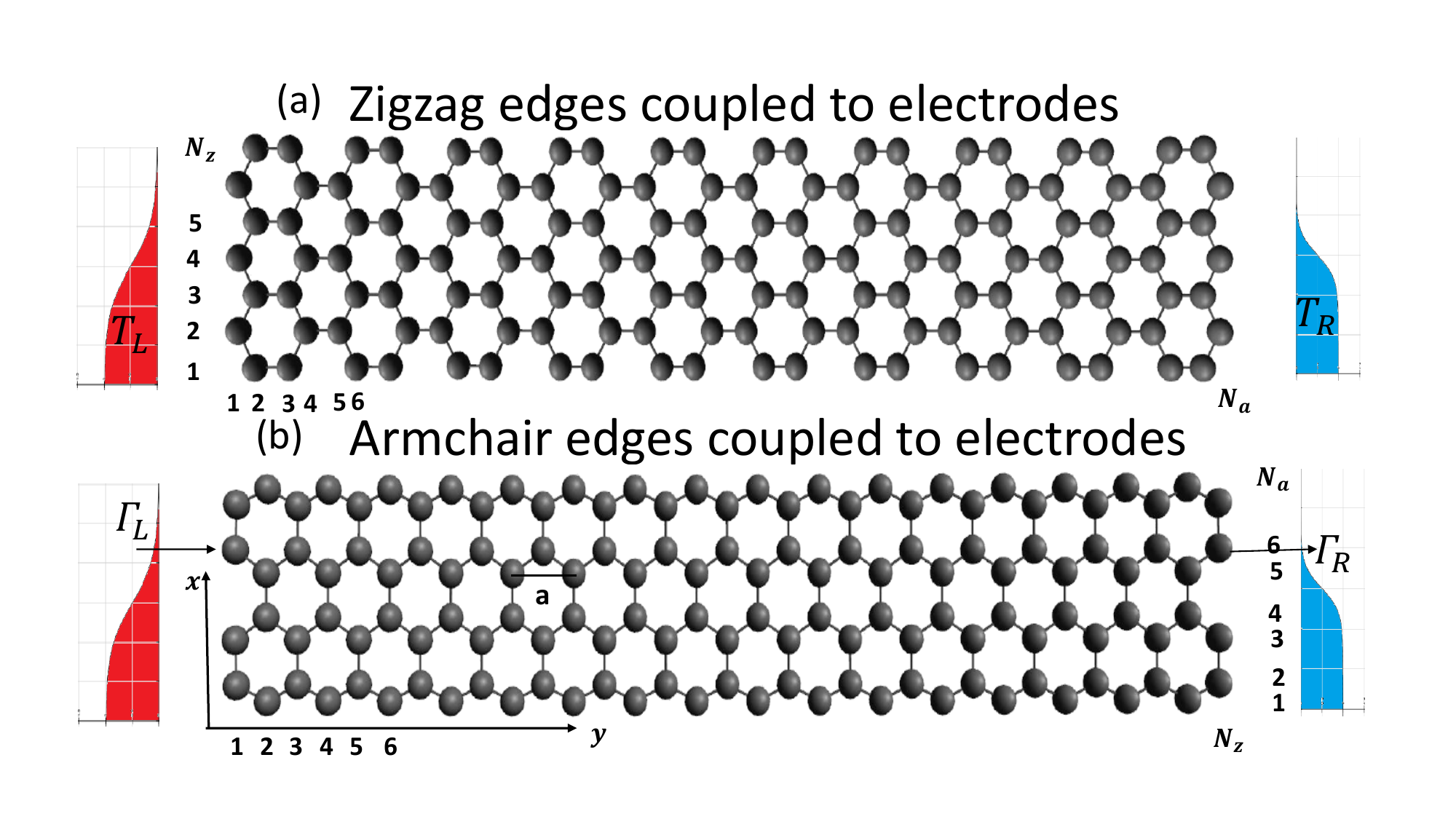}
\caption{Schematic diagram of a finite graphene coupled to
metallic electrodes. $\Gamma_{L}$ ($\Gamma_R$) denotes the
tunneling rate of the electrons between the left (right) electrode
and the leftmost (rightmost) atoms of a finite graphene
nanoribbon. $T_L$ and $T_R$ denote the temperature of the left and
the right electrodes, respectively. (a) and (b) show the geometry
with the zigzag and armchair edges coupled to the electrodes,
respectively. The lattice constant of graphene is $a=2.46 \AA$.}
\end{figure}

\section{Formalism}
The hexagonal crystal structure of graphene results from the
$\sigma$ bonding formed by $sp^2$ hybridized orbitals, while the
$p_z$ orbitals form the $\pi$ bands that play an important role
for electron transport near Fermi energy.$^{30-36)}$ Similar to
benzene,$^{39)}$ it is expected that the $\sigma$ bands of
graphene are well separated from the $\pi$ bands. Therefore, it is
a good approximation to employ a tight-binding model with one
$p_z$ orbital per atomic site to describe the electronic states
near the Fermi level of a GNR .$^{31)}$ To model the
thermoelectric properties of a GNR connected to the electrodes,
the Hamiltonian of the system depicted in Fig.~1(a) is written as
$H=H_0+H_{GNR}$,$^{40)}$ where
\begin{small}
\begin{eqnarray}
H_0& = &\sum_{k} \epsilon_k a^{\dagger}_{k}a_{k}+
\sum_{k} \epsilon_k b^{\dagger}_{k}b_{k}\\
\nonumber &+&\sum_{\ell}\sum_{k}
V^L_{k,\ell,j}d^{\dagger}_{\ell,j}a_{k}
+\sum_{\ell}\sum_{k}V^R_{k,\ell,j}d^{\dagger}_{\ell,j}b_{k} + h.c.
\end{eqnarray}
\end{small}
The first two terms of Eq.~(1) describe the free electrons in the
left and right metallic electrodes. $a^{\dagger}_{k}$
($b^{\dagger}_{k}$) creates  an electron of with momentum $k$ and
energy $\epsilon_k$ in the left (right) electrode.
$V^L_{k,\ell,j=1}$ ($V^R_{k,\ell,j=N_a}$) describes the coupling
between the left (right) lead with its adjacent atom in the
$\ell$-th row.
\begin{small}
\begin{eqnarray}
H_{GNR}&= &\sum_{\ell,j} E_{\ell,j}
d^{\dagger}_{\ell,j}d_{\ell,j}\\ \nonumber&+&
\sum_{\ell,j}\sum_{\ell',j'} t_{(\ell,j),(\ell', j')}
d^{\dagger}_{\ell,j} d_{\ell',j'} + h.c,
\end{eqnarray}
\end{small}
where { $E_{\ell,j}$} is the on-site energy for the $p_z$ orbital
in the ${\ell}$-th row and $j$-th column. Here, the spin-orbit
interaction is neglected. $d^{\dagger}_{\ell,j} (d_{\ell,j})$
creates (destroys) one electron at the atom site labeled by
($\ell$,$j$) where $\ell$ and $j$, respectively are the row and
column indices as illustrated in Fig.~1. $t_{(\ell,j),(\ell',
j')}$ describes the electron hopping energy from site ($\ell$,$j$)
to site ($\ell'$,$j'$). The electron wave functions of the zigzag
edge states of GNRs are well localized,$^{30-32)}$ and the Coulomb
repulsion between two zigzag edge-state electrons can be strong
when they are close. Thus, the Coulomb repulsion effect on
electron transport through the edge states can be significant when
the average occupancy of each site is larger than 0.5 in steady
state.$^{41)}$ On the other hand, the wave functions of the
electrons in the bulk-like states are delocalized; hence their
weak electron Coulomb interactions can be neglected. Because the
transport behavior of zigzag edge states behaves like that of
coupled quantum dots (to be demonstrated in the next section), its
behavior can be analyzed by considering the Coulomb blockade
effect in coupled quantum dots with orbital or valley degeneracy
.$^{41)}$ The tight-binding parameters used for GNR is
$E_{\ell,j}=0$ and $t_{(\ell,j),(\ell',j')}=t_{pp\pi}=-2.7$ eV for
nearest-neighbor hopping only.

To study the transport properties of a GNR junction connected to
electrodes, it is convenient to use the Keldysh-Green's function
technique.$^{40,42)}$ Electron and heat currents leaving the
electrodes can be expressed as
\begin{eqnarray}
J &=&\frac{g_{s}e}{h}\int {d\varepsilon}~ {\cal
T}_{LR}(\varepsilon)[f_L(\varepsilon)-f_R(\varepsilon)],
\end{eqnarray}
and
\begin{equation}
% & &Q_{e,L(R)}\\ &=&\frac{\pm g_{s}}{h}\int {d\varepsilon}~ {\cal T}_{LR}(\varepsilon)(\varepsilon-\mu_{L(R)})[f_L(\varepsilon)-f_R(\varepsilon)]\nonumber
Q_{e,L(R)}=\frac{\pm g_{s}}{h}\int {d\varepsilon}~ {\cal T}_{LR}(\varepsilon)(\varepsilon-\mu_{L(R)})[f_L(\varepsilon)-f_R(\varepsilon)]
\end{equation}
where $g_{s}=2$ denotes the spin degeneracy.
$f_{\alpha}(\varepsilon)=1/\{\exp[(\varepsilon-\mu_{\alpha})/k_BT_{\alpha}]+1\}$
denotes the Fermi distribution function for the $\alpha$-th
electrode, where $\mu_\alpha$  and $T_{\alpha}$ are the chemical
potential and the temperature of the $\alpha$ electrode. $e$, $h$,
and $k_B$ denote the electron charge, the Planck's constant, and
the Boltzmann constant, respectively. ${\cal T}_{LR}(\varepsilon)$
denotes the transmission coefficient of a $GNR$ connected to
electrodes, which can be solved by the formula {$ {\cal
T}_{LR}(\varepsilon)=4Tr[{\Gamma}_{L}(\varepsilon){G}^{r}(\varepsilon){\Gamma}_{R}(\varepsilon){G}^{a}(\varepsilon)]$}
,$^{43,44)}$ where (${\Gamma}_{L}(\varepsilon)$ and
${\Gamma}_{R}(\varepsilon)$) denote the tunneling rate at the left
and right leads, and {${G}^{r}(\varepsilon)$ and
${G}^{a}(\varepsilon)$ are the retarded and advanced Green's
function of the GNR.} Note that  both the Green's functions and
the tunneling rates in the transmission coefficient ${\cal
T}_{LR}(\varepsilon)$ of Eqs. (3) and (4) are matrices.
$\Gamma_{\alpha}(\varepsilon)=-Im(\Sigma^r_{\alpha}(\varepsilon))$
result from the imaginary part of self energies determined by
$V^{L}_{k,\ell,j=1}$ and $V^{R}_{k,\ell,j=N_a}$ (see Fig. 1(a)).
Therefore, the matrix form of $\Gamma_L(\varepsilon)$ is always
different from that of $\Gamma_R(\varepsilon)$. For metals such as
gold, the density of states is approximately constant near the
Fermi energy such that the wide-band limit is a good
approximation.$^{43)}$ In the wide-band limit, the
$\Gamma_{L(R)}(\varepsilon)$ are replaced by constant matrices
$\Gamma_{L(R)}$. However, only the diagonal entries are non-zero.
In the following, we choose a symmetric coupling such that the
non-vanishing matrix elements of $\Gamma_{\alpha}$ all take the
same value $\gamma_L=\gamma_R=\gamma$.$^{43)}$ In the current
study, only integers of $N_a/4$ are considered here. For the case
with odd numbers of $N_a$, one edge of the GNR will have broken
bonds which lead to unwanted dangling-bond states.

In the linear response regime, the electrical conductance ($G_e$),
Seebeck coefficient ($S$) and electron thermal conductance
($\kappa_e$) are given by $G_e=e^2{\cal L}_{0}$, $S=-{\cal
L}_{1}/(eT{\cal L}_{0})$ and $\kappa_e=\frac{1}{T}({\cal
L}_2-{\cal L}^2_1/{\cal L}_0)$ with ${\cal L}_n$ ($=0,1,2$)
defined as
\begin{equation}
{\cal L}_n=\frac{2}{h}\int d\varepsilon~ {\cal
T}_{LR}(\varepsilon)(\varepsilon-\mu)^n\frac{\partial
f(\varepsilon)}{\partial \mu}.
\end{equation}
Here $f(\varepsilon)=1/(exp^{(\varepsilon-\mu)/k_BT}+1)$ is the
Fermi distribution function of electrodes at equilibrium
temperature $T$ and chemical potential $\mu$. As for the phonon
contribution to the thermal conductivity, $\kappa_{ph}$, we adopt
the calculated results reported in ref.~[45], where the phonon
scattering from defects in carbon nanotubes is studied
theoretically. As can be seen from Eq. (5), the transmission
coefficient, ${\cal T}_{LR}(\varepsilon)$  plays a significant
role for electron transport and thermoelectric properties.

\section{Results and discussion}

\subsection{Zigzag edges coupled to the electrodes}
\subsubsection{ Graphene nanoribbons with $N_a \gg N_z$ }

Armchair graphene nanoribbons (AGNRs) can be either metallic (when
$N_z=3m+2$) or semiconducting (when $N_z=3m$ or $3m+1$).$^{30)}$
Here $m$ is an integer. For reference, the calculated electronic
band structures of GNRs with infinite length are given in appendix
A. For semiconducting cases, the size of the gap  is inversely
proportional to $N_z$ (or the width of the AGNR).$^{31)}$ To
examine the thermoelectric properties of GNRs, we have calculated
$G_e$, $S$ {and power factor $PF$} as functions of $\mu$ for
different $k_BT$ with $N_a=100$ (length $L_a=10.5$nm) and $N_z=24$
(length $L_z=2.8$nm) in Fig. 2. Due to the quantum confinement
arising from finite $N_a$ and $N_z$, discrete peaks show up in the
$G_e$ spectra. Because the zigzag edge states are localized
electronic states that decay exponentially toward the center of
the ribbon (see charge density in appendix B. 1), the probability
of electrons transport between the electrodes tunneling through
zigzag edge states is vanishingly small for a long channel length
($L_a=10.5$nm).

As shown in Fig. 2(a) the GNR has an energy gap around $0.5~eV$
for $N_z=24$ and $N_a=100$, which is slightly larger than that of
an infinitely-long AGNR with $N_z=24$. The $G_e$ peak shows a
thermal broadening behavior with respect to temperature.
Oscillatory behavior of $G_e$ can be observed even at
$k_BT=27meV$. The temperature-dependent maximum Seebeck
coefficient ($|S_{h,max}|$ for hole or $|S_{e,max}|$ for electron)
shows an impressive value inside the gap region$^{29)}$. Note that
the unit $(k_B/e)$ equals to $86.25\mu V/K$. We have
{$|S_{h,max}|=|S_{e,max}|$}=$431 \mu V/K$ for $k_BT=27$ meV. As
seen in Fig. 2, due to the electron-hole symmetry $G_e(\mu)$ is
symmetric while $S(\mu)$ is antisymmetric with respect to the sign
change of $\mu$. Thus, $S$ vanishes at $\mu=0$ at any temperature.
In Fig. 2, $G_e$ is smaller than the quantum conductance,
$G_0=2e^2/h$. For thermoelectric devices it is important to
optimize the electrical power outputs as well as efficiency. To
increase power factor $PF=S^2G_e$, one can increase $G_e$ by
increasing $N_z$ or tunneling rates. Thus, we consider case
$N_z\gg N_a$ below.

\begin{figure}[h]
\centering
\includegraphics[angle=0,scale=0.3]{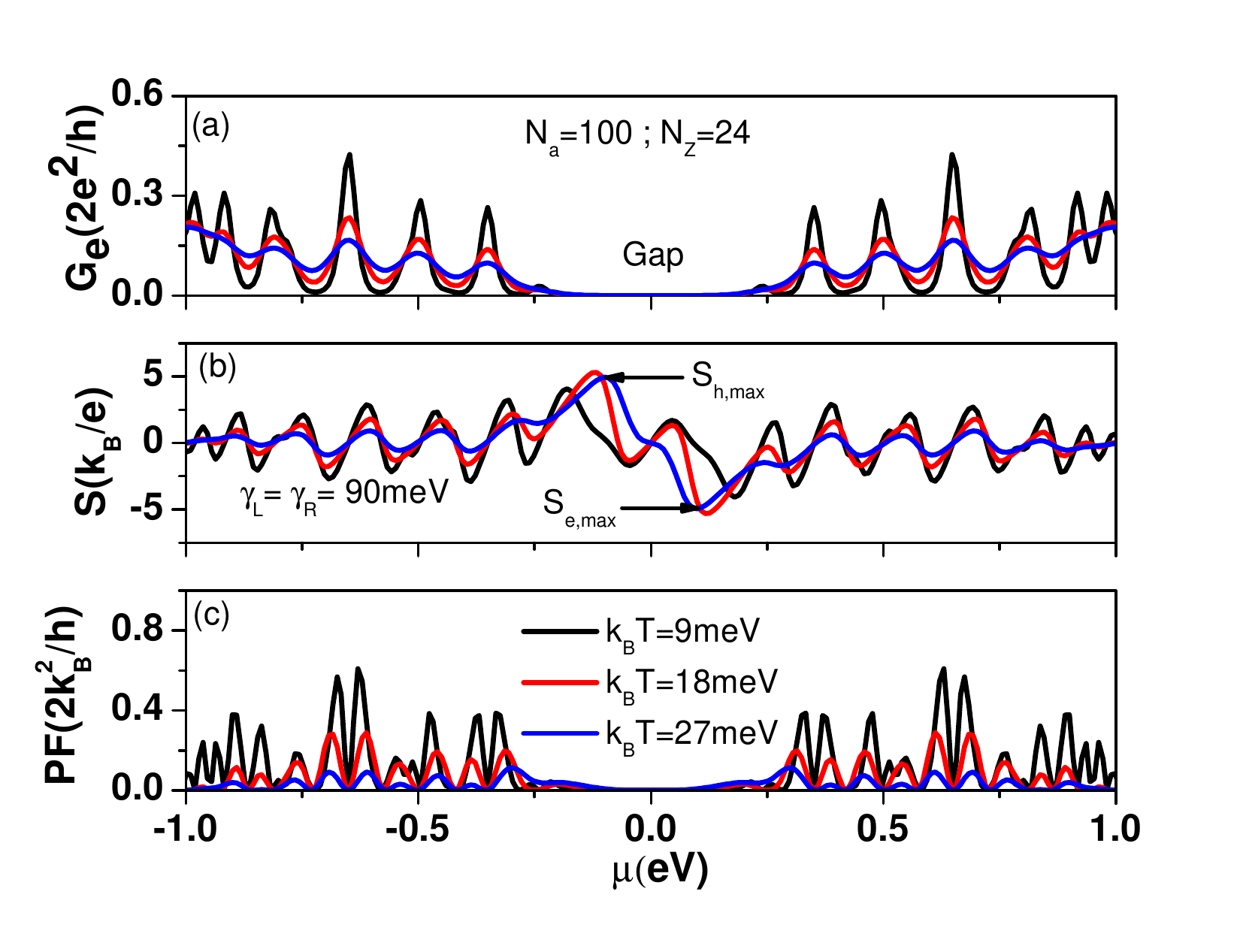}
\caption{(a) Electrical conductance $G_e$,(b) Seebeck coefficient
$S$ and (c) power factor $PF$ as functions of $\mu$ for different
temperatures at $N_a=100$ ($L_a=10.5$ nm) and
$N_z=24$($L_z=2.8$nm). We have adopted electron tunneling rates
$\gamma_L=\gamma_R=90$ meV.}
\end{figure}

\subsubsection{ Graphene nanoribbons with $N_z \gg N_a$}

Figure 3 shows the calculated conductance ($G_e$), Seebeck
coefficient ($S$), and power factor ($PF$) of a GNR with $N_z=99$
($L_z=12$nm) and $N_a=20$($L_a=1.988$nm) as functions of $\mu$ for
various temperatures. According to Fig. A.2, only zigzag edge
states provide tunneling channels between the electrodes when
$|\epsilon| \le 1.0 eV $. Here, $G_e$ is much enhanced because
more carbon atoms of zigzag edges are in contact with the
electrodes in this arrangement. The peak at $\mu=0$ (marked by
$\Sigma_0$) arises from the zigzag edge states. Because zigzag
edge states are localized, the width of the $\Sigma_0$ peak is
very sensitive to the $N_a$ value. In addition, the broadening of
$\Sigma_0$ peak also depends on the tunneling rates, $\gamma_L$
and $\gamma_R$. Here we set $\gamma_L=\gamma_R=90$ meV. The
quantum confinement effect is clearly seen for $|\mu| \ge 0.2eV$.
In Fig.~3(b) {the maximum Seebeck coefficients are much smaller
than $S_{e(h),max}$} shown in Fig. 2(b). The behavior of Seebeck
coefficient shown in Fig. 3(b) can be described roughly by
$S\approx -\frac{k^2_BT}{e}\frac{1}{G_e(\mu,T)}\frac{\partial
G_e(\mu,T)}{\partial \mu}$. This implies that the value of $S$
becomes very small when $G_e$ is insensitive to $\mu$. Fig.~3(c)
shows that GNRs give a significant power factor at $k_BT=27meV$ as
$\mu$ is tuned away from 0. However, due to large $\kappa_e$ (not
shown here), the thermoelectric efficiency of GNRs in the
situation of $N_z >> N_a$ is suppressed.

\begin{figure}[h]
\centering
\includegraphics[angle=0,scale=0.3]{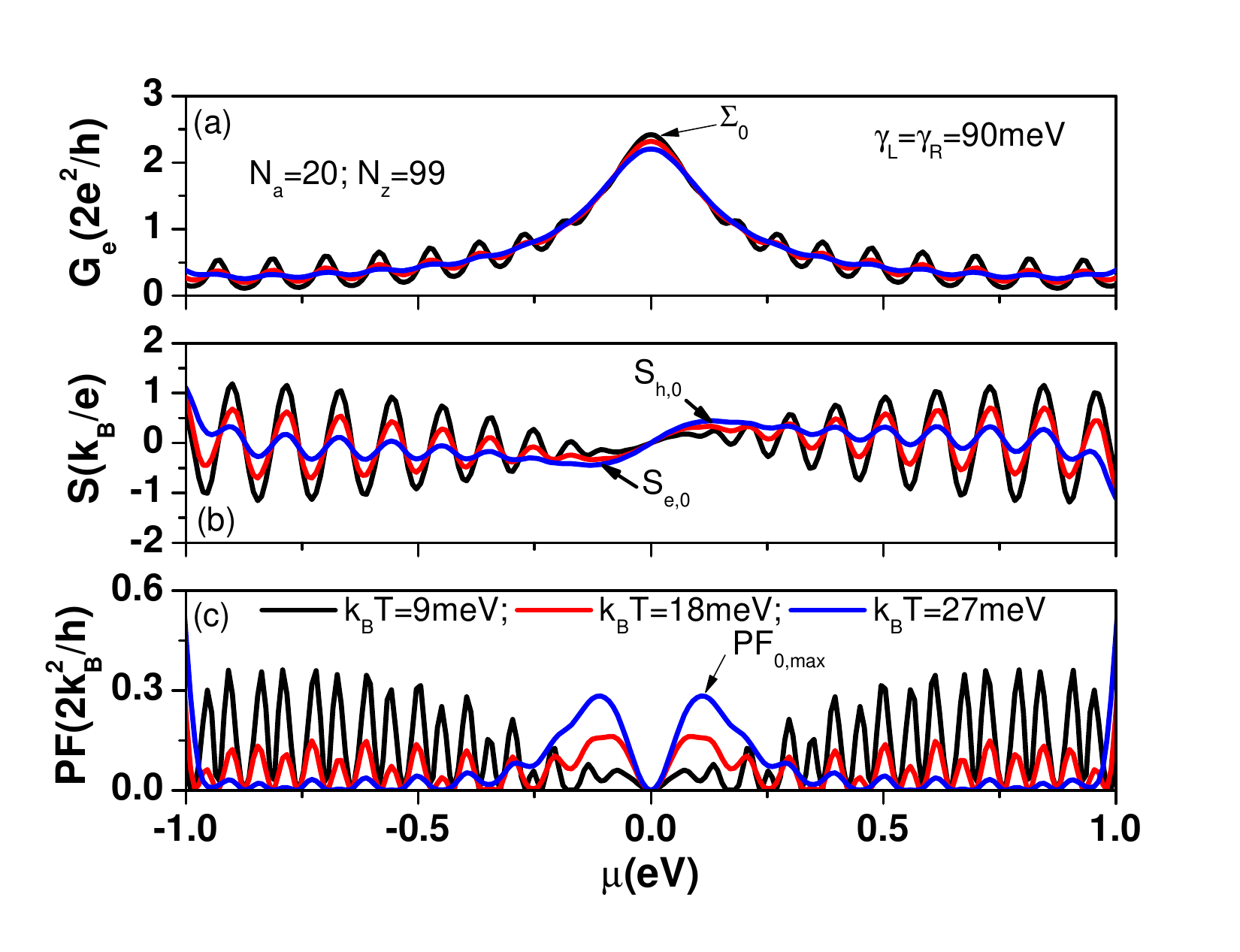}
\caption{ (a)Electrical conductance $G_e$,(b) Seebeck coefficient
$S$ and (c) power factor $PF$ as functions of $\mu$ for different
temperature values at $N_a=20$ ($L_a=1.988$nm) and $N_z=99$
($L_z=12$nm). $\gamma_L=\gamma_R=90$ meV.}
\end{figure}

\subsection{Armchair edges coupled to the electrodes}

To further understand the zigzag edge effect on the thermoelectric
properties of GNRs, we next consider a GNR  with armchair edges
coupled to the metallic electrodes (as illustrated in Fig.~1(b)).

\subsubsection{ Graphene nanoribbons with $N_z \gg N_a$}

In contrast to the situation of Figs. 2 and 3, here we consider
the scenario with armchair edges connected to the electrodes.
Fig.~4 shows electrical conductance $G_e$, Seebeck coefficient
$S$, power factor $PF$ and figure of merit $ZT_e$ as functions of
$\mu$ for various temperatures at $N_a=24$ ($L_a=2.41$nm) and
$N_z=121$ ($L_z=14.76$nm). Here we have adopted
$\gamma_{L(R)}=9meV$ to reduce the broadening effect on $G_e$
spectra.  Since $N_z\gg N_a$ in this case, the electronic states
are comparable to the case of an infinitely-long ZGNR with
$N_a=24$ (as shown in Fig. A.2). We note that strictly speaking
there is always an energy splitting of the pair of zigzag-edge
states for ZGNR with finite width, although the splitting can be
quite tiny for those edge states with $|k|>\frac{2\pi}{3a}$. Thus,
at zero temperature, we find $G_{e}(\mu=0)=0$. However, at finite
temperature with $k_BT$ larger than the energy splitting mentioned
above, the $\Sigma_0$ peak in the $G_e$ spectrum will appear to
have a maximum at $\mu=0$ due to temperature-smearing effect. When
$\mu$ is tuned away from the zero-energy mode into the fork-shaped
region (see Fig. A. 2), the $G_e$ spectra show a high density of
peaks (indicated by $\Sigma_1$) with nearly uniform height, where
the spacing between consecutive peaks is approximately $90$meV. As
seen in Fig. 4(a), $G_e$ is highly enhanced near $|\mu|= 0.9eV$.
Such an enhancement of $G_e$ is attributed to the combination of
zigzag-edge states and the second subband. The fast oscillation of
$G_e$ spectra resulting from quantum confinement give rise to
significant $S$ values, which are suppressed as temperature
increases. In addition, $S$ is significantly suppressed once
$|\mu|$ increases beyond $0.9 eV$ (the onset of the second
subband). The maximum power factor occurs near the onset of the
second subband band. Remarkable $ZT_e$ values are observed at
$|\mu|$ near $0.9eV$ in Fig. 4(d). Here, we have omitted
$\kappa_{ph}$ in the calculation of $ZT_e$. This implies that
finding a mechanism to reduce  $\kappa_{ph}$ to a value below
$\kappa_{e}$ is important in the realization of high-efficiency
thermoelectric materials.$^{2)}$

\begin{figure}[h]
\centering
\includegraphics[angle=0,scale=0.3]{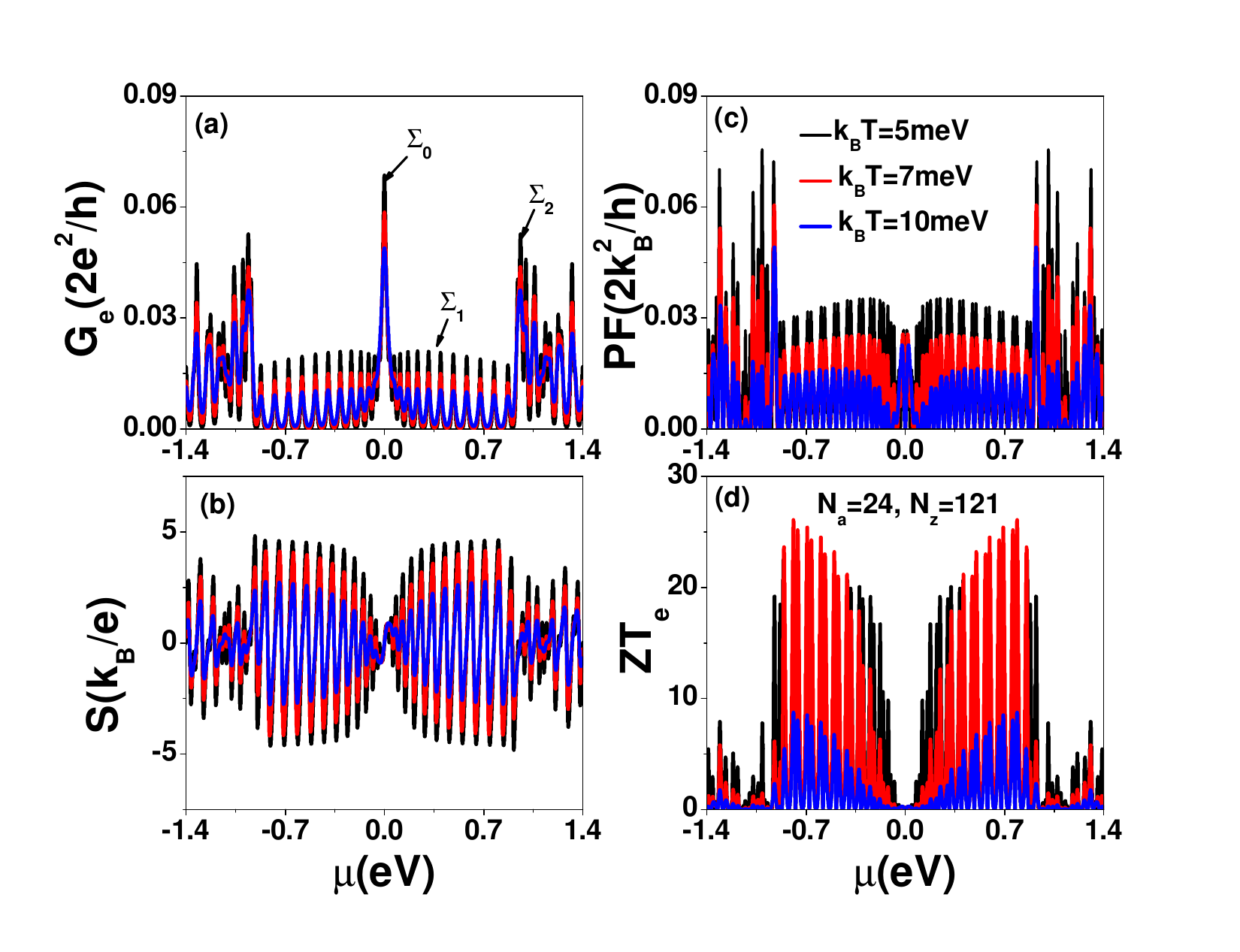}
\caption{(a) Electrical conductance $G_e$, (b) Seebeck coefficient
$S$, (c) power factor $PF$ and (d) {figure of merit, $ZT_e$} as
functions of $\mu$ for GNRs with $N_z=121$ ($L_z=14.76$nm) and
$N_a=24$ ($L_a=2.41$nm). $\gamma_L=\gamma_R=9$ meV.}
\end{figure}

\subsubsection{Graphene nanoribbons with $N_a \gg N_z$}
Unlike $G_e$ (which prefers band-like situation), the best $S$
comes from {discrete} electronic states (atomic-like situation).
Therefore, we consider the situation of $N_a \gg N_z$ to make the
$\Sigma_0$ peak (due to edge states) well separated from the
higher subband states. Figure 5 shows $G_e$ as a function of $\mu$
for GNRs with $N_a=100$ and $N_z=10, 15$, and $20$ (corresponding
to $L_z=1.1, 1.72 $ and $2.34$ nm), at zero temperature. Here, the
GNRs with $N_z=10$, and $15$ are semiconducting, while the GNR
with $N_z=20$ is metallic. For the semiconducting phase, a sizable
gap opens up between the conduction band and valence band. Most
importantly, the zigzag edge state appears at the mid gap (the
peak marked $\Sigma_0$) as can be seen in Fig. 5(a,b). We note
that there are two zigzag edge states localized at the top and
bottom ends of the GNR. The charge density of GNRs provided in
Fig.~B.1 reveal that these zero-energy modes are derived from
zigzag edge states. Obviously, the variation of $N_z$ not only
changes the phase of GNRs but also the magnitude and width of
$\Sigma_0$. Such a phenomenon does not exist in infinite long
AGNRs.$^{46,47)}$

\begin{figure}[h]
\centering
\includegraphics[angle=0,scale=0.3]{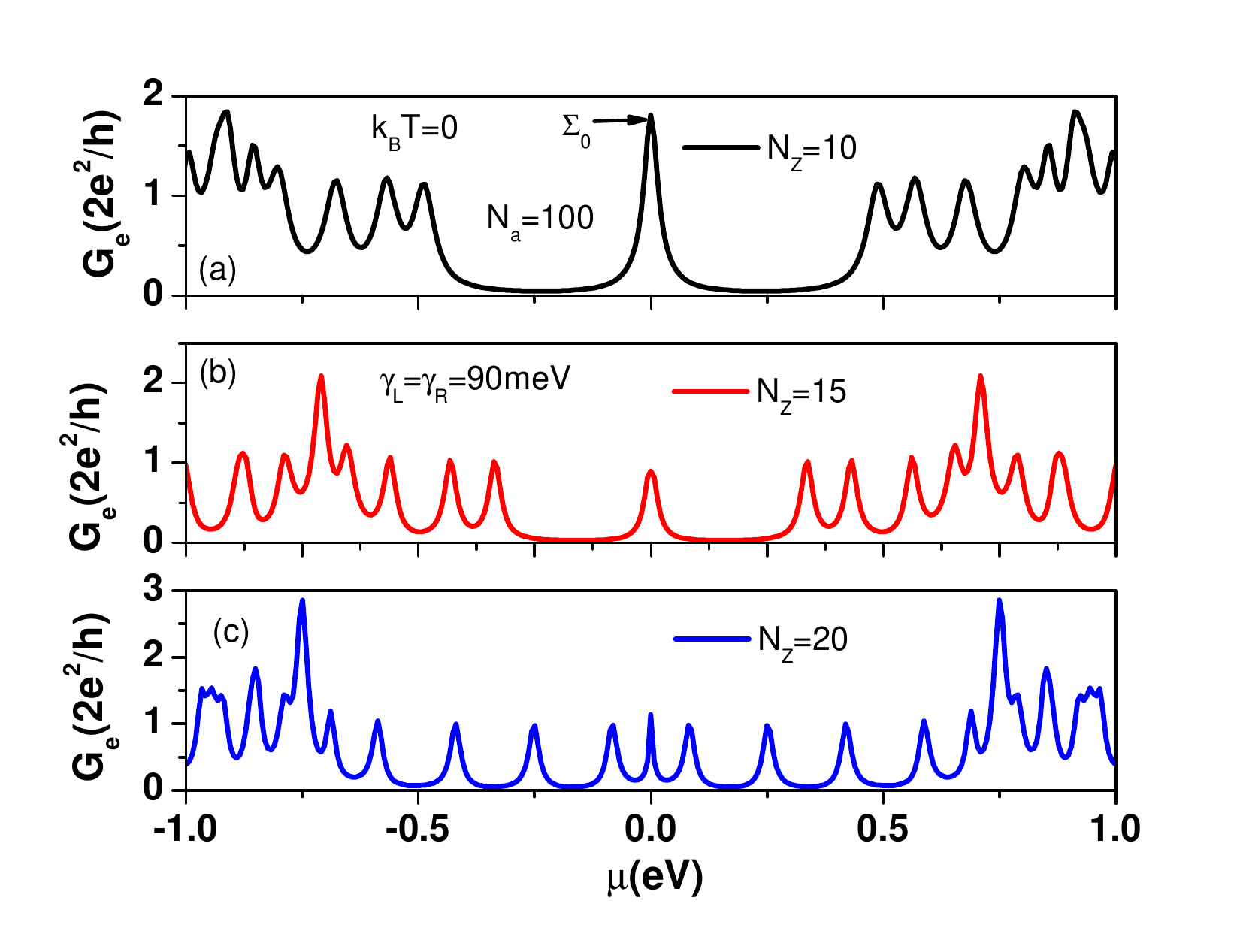}
\caption{Electrical conductance as functions of $\mu$ for various
$N_z$ numbers at $k_BT=0$ and $N_a=100$.$\gamma_L=\gamma_R=90$
meV.}
\end{figure}

Mahan and Sofo proposed to employ a single quantum dot (QD) to
realize a Carnot heat engine.$^{48)}$ The $G_e$ spectra near
$\mu=0$ shown in Fig. 5(a,b) are similar to the $G_e$ spectra of
nanoscale semiconductor QDs.  To further clarify the effect of
zero-energy modes on thermoelectric coefficients, we show the
calculated {$G_e$, $S$, $PF$ and $\kappa_e$} as functions of $\mu$
for the case with $N_z=15$ (Fig.~5(b)) at $k_BT=9$meV in Fig.~6.
The electrical conductance due to zigzag edge states ($\Sigma_0$)
is suppressed with increasing temperature. Such a behavior is very
common in a single QD system. The maximum Seebeck coefficient
$S_{h(e),0,max}$ resulting from $\Sigma_0$ depends on the gap
around the $\Sigma_0$ peak ($\Delta=E_c-E_v$) (or see Fig. 7).
Although the maximum $PF_{0,max}$ arising from zero energy modes
is slightly smaller than $PF_{B,max}$ resulting from bulk states,
its electron thermal conductance could be very small. This
indicates that the thermoelectric efficiency of zero energy modes
is better than that of bulk states.

%Comparing with Fig. 4(d), we found that the $\kappa_e$ derived
%from zero-energy modes of AGNRs are much smaller than that of
%ZGNRs.

\begin{figure}[h]
\centering
\includegraphics[angle=0,scale=0.3]{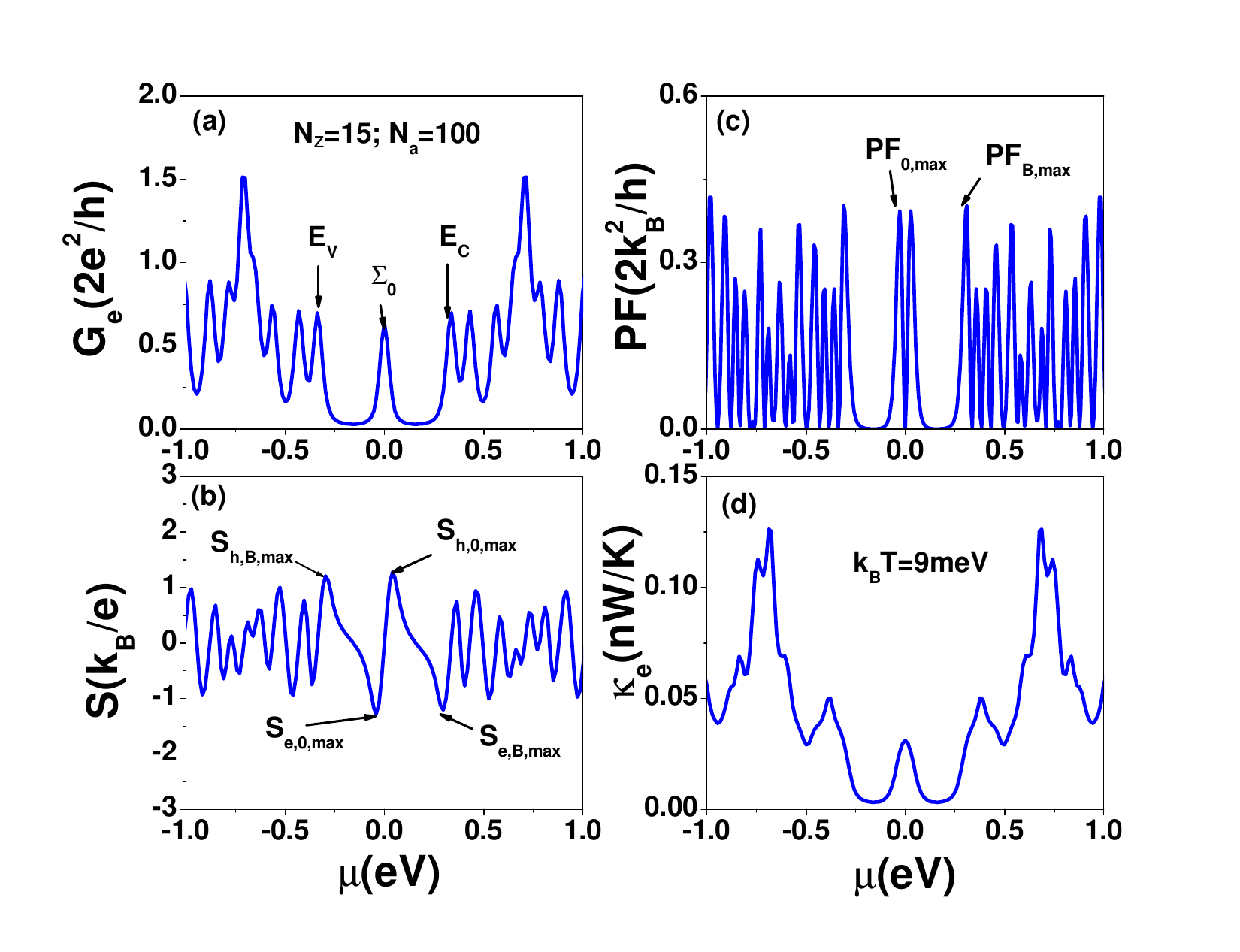}
\caption{(a) Electrical conductance $G_e$, (b) Seeback coefficient
$S$, (c) power factor $PF$ and (d) electron thermal conductance
$\kappa_e$ as functions of $\mu$ for various temperatures at
$N_a=100$,and $N_z=15$. $\gamma_L=\gamma_R=90$ meV.}
\end{figure}

Next we examine the size effect of $N_z$ on thermoelectric
quantities. We show the calculated $G_e$, $S$, $PF$ and the figure
of merit ($ZT_e=S^2G_eT/\kappa_e$) as functions of $\mu$ for
different values of $N_z$ at $k_BT=27$ meV (near room temperature)
for the case with a smaller tunneling rate ($\gamma_L=\gamma_L=9$
meV) in Fig. 7. We kept $N_z=3m$ ($m$ is an integer) in Fig.~7 to
maintain a finite band gap for GNRs. As seen in Fig.~7(a), the
electrical conductance $\Sigma_0$ is reduced with increasing $N_z$
{(also see Fig.5)}. In addition, the maximum $S_0$ value is
degraded with the increase of $N_z$ since the nearest peaks get
close to $\Sigma_0$. It is worth noting that the behavior of
$S_{0}=\mu/(eT)$ is observed in Fig. 7(b). Such a feature was
theoretically reported in our previous study.$^{49)}$ Meanwhile,
the maximum $PF$ and $ZT$ occur at the condition with
{$|\mu|/(k_BT)=2.4$}. In the calculation of $ZT_e$, we have
considered $\kappa_{ph}=0$ to estimate the maximum $ZT$ values. If
instead we adopted $\kappa_{ph}=\frac{\pi^2k^2_B
T}{3h}$,$^{28,45)}$ $ZT$ will reduce significantly. In Fig. 7, we
have adopted $\gamma_{L(R)}=9meV$, which is one order of magnitude
smaller in comparison with cases considered in figures(5) and (6).
Tunneling rates could be affected by the Schottky barrier contact
between metal and semiconductor.$^{50)}$

It will be a big challenge to make $\kappa_{ph}$ much smaller than
$\kappa_{e}$ in the situation of $N_a\gg N_z$. Due to a short
channel length ($L_z=1.72$nm), both electrons and phonons remain
in the ballistic transport regime. Nevertheless, the zero-energy
modes resulting from zigzag edge states are very robust. The
transport of zero energy modes of GNRs are topologically protected
against scattering while phonons are significantly scattered when
defects or disorders are introduced into the transport system. As
a consequence, $\kappa_{ph}$ is highly reduced.$^{51)}$ So far, a
quantitative study on how defects influence the electron transport
through zigzag-edge states in finite-size GNRs is still
lacking.$^{52)}$

\begin{figure}[h]
\centering
\includegraphics[angle=0,scale=0.3]{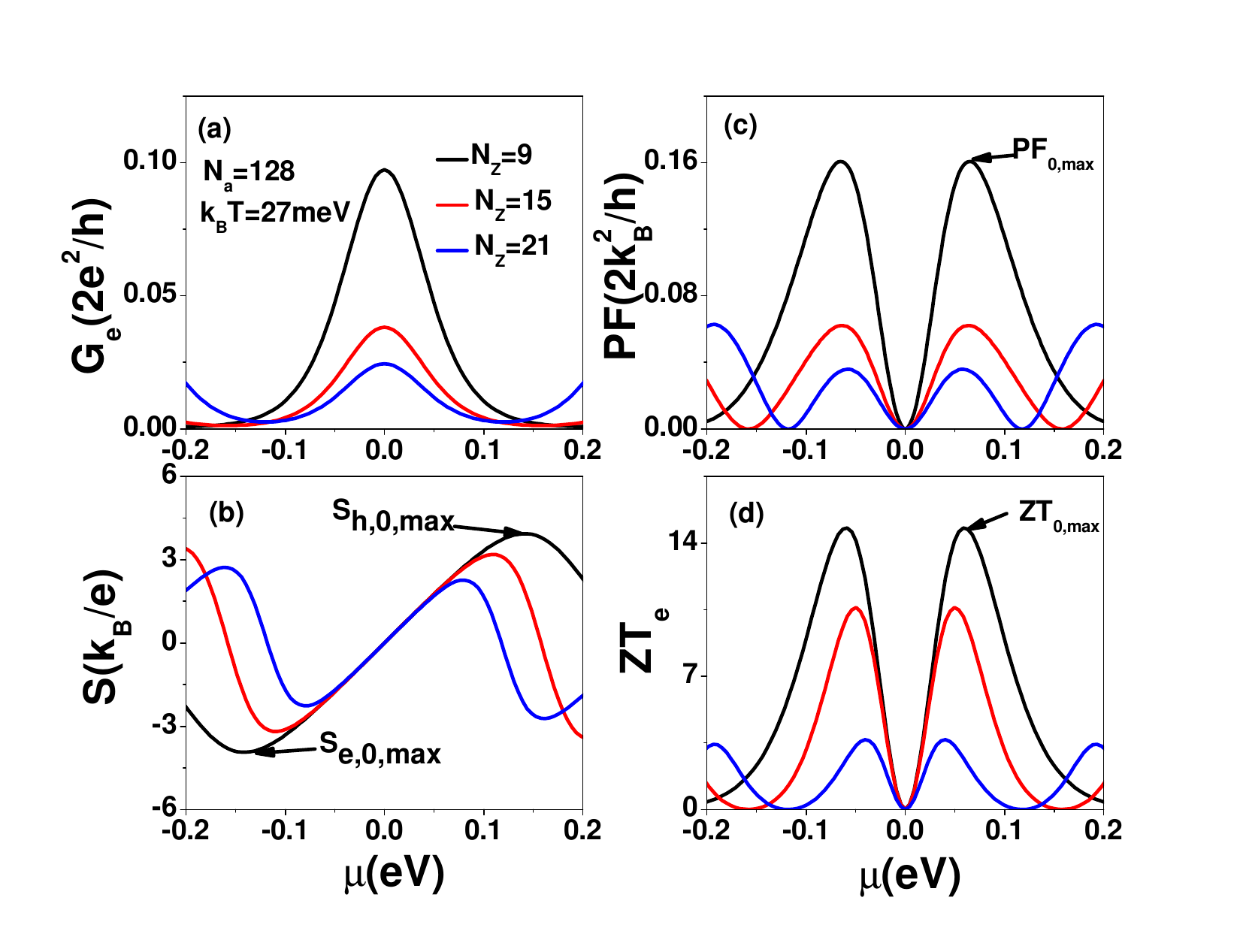}
\caption{(a) Electrical conductance, (b) Seebeck coefficient, (c)
power factor and (d) figure of merit ($ZT_e$) as functions of
chemical potential for different $N_z$ values at $N_a=128$
($L_a=13.49$nm), $k_BT=27 meV$ and $\gamma_L=\gamma_R=9meV$. }
\end{figure}

To examine how robust the $\Sigma_0$ peak resulting from zigzag
edge states is against the presence of defects,$^{51,52)}$ we show
in Fig.~8 the effect on $G_e$ due to atomic vacancies randomly
distributed in GNRs. We plot $G_e$ as a functions of $\mu$ for
different defect locations at $k_BT=27meV$ and
$\gamma_{L(R)}=27meV$ for finite-size GNRs with $N_z=9$
($L_z=0.98$nm) and $N_a=128$ ($L_a=13.49$nm). We use the concept
of orbital removal (by setting the energy level of the defect site
to at a large value $E_d=1000 eV$) to mimic the creation of a
vacancy. Fig.~8(a) shows that when the vacancy occurs at any
location ($\ell,j$) away from the zigzag edge (with $\ell > 1$ or
$\ell < N_a$) $G_e$ of the GNR is almost the same as the
defect-free (DF) case, where the peak height of $G_e$ is close to
$0.24 G_0$. On the other hand, when a vacancy occurs on one zigzag
edge (with $\ell=1$ or $\ell=N_a$) the conductance $G_e$ reduces
by 1/2, indicating the contribution to the electron transport by
the zero-energy mode at that edge is blocked by the presence of
vacancy, the contribution due the other edge remains intact. As
shown in Fig. 8(b), the effect of defects on the behavior of power
factor is quite similar to that of $G_e$. This indicates that
Seebeck coefficient, $S_0$ resulting from zigzag-edge states is
essentially unchanged against defect scattering. This means that
$S_0$ will not reduce appreciably even though a vacancy blocks the
transport on one zigzag edge. Vacancies away from the zigzag edge
won't influence the $\Sigma_0$ peak and $S_0$ spectrum, but they
are expected to reduce $\kappa_{ph}$ significantly, thus improving
the figure of merit $ZT$.$^{51)}$

\begin{figure}[h]
\centering
\includegraphics[angle=0,scale=0.3]{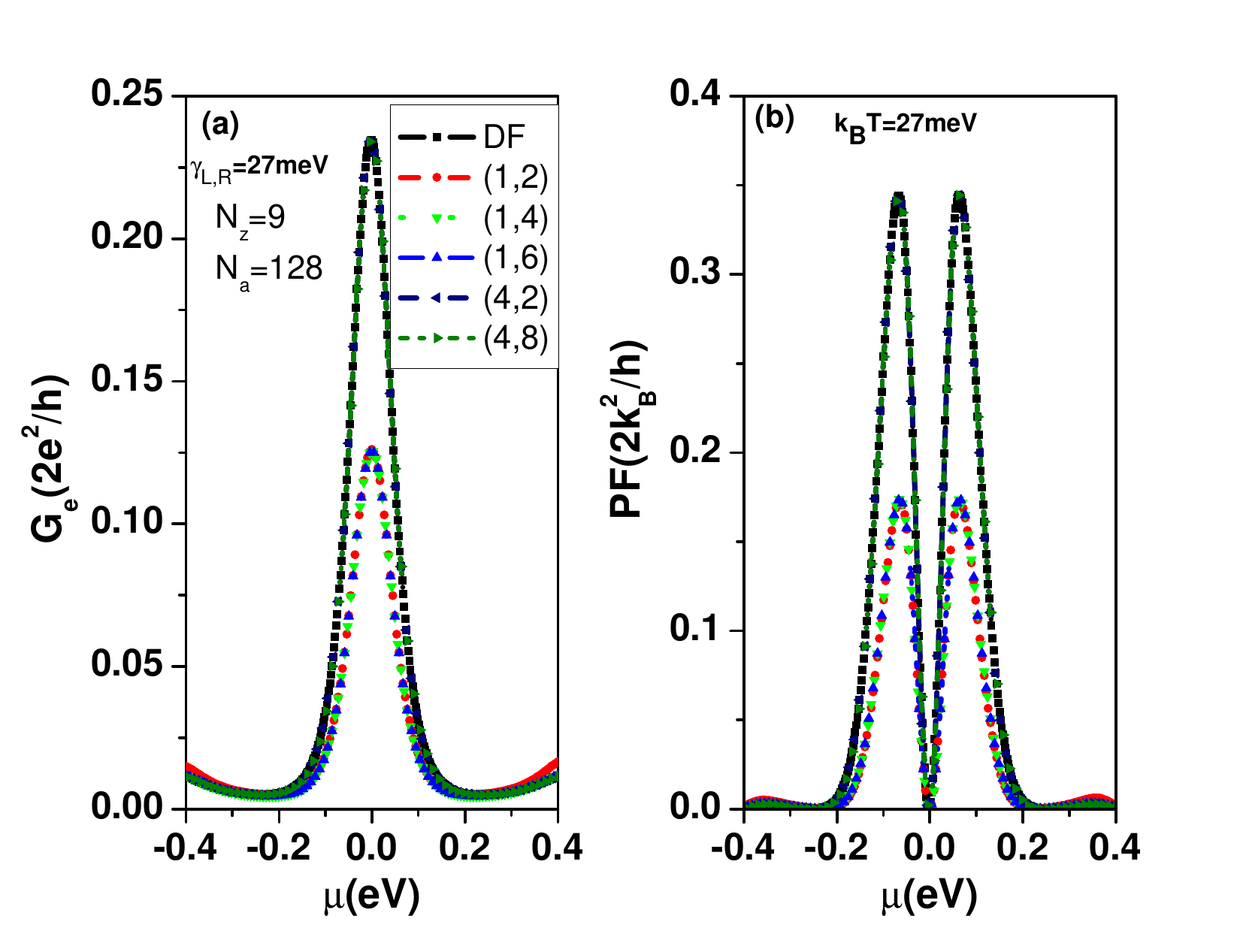}
\caption{(a)Electrical conductance and (b) power factor as
functions of chemical potential for different defect locations at
$k_BT=27meV$, $\gamma_{L(R)}=27meV$, $N_z=9$ and $N_a=128$
($L_a=13.49$nm).}
\end{figure}

The contact problem between metal and semiconductor plays a
remarkable role in the novel applications of 2D electronics
.$^{50)}$ Here, we show that the tunneling rate ($\gamma_{L,R}$),
which depends on the contact property, can significantly affects
the thermoelectric properties of GNRs. The calculated electrical
conductance, Seebeck coefficient, power factor and figure of merit
as functions of chemical potential  for various tunneling rate
$\gamma_L=\gamma_R$ at $k_BT=27$ meV are shown in Fig.~9(a)-(d).
From results of Fig. 9(a), we see that it is not easy to measure
$\Sigma_0$ peak when $\gamma_{L,R}/(2k_BT)\ll 1$. In contrast to
$G_e$, $S_0$ prefers weaker coupling strength between metal and
GNR. Although $S_0$ can be improved by reducing $\gamma_{L,R}$,
the electrical power output of GNRs is greatly enhanced as
$\gamma_{L,R}$ increases. Combining the above considerations we
found that the maximum $ZT$ occurs when the tunneling rate matches
the operating temperature, i.e. $\gamma_{L,R}=k_BT$, when the
phonon thermal conductivity is included. Here, we have adopted
$\kappa_{ph}=F_s \frac{\pi^2k^2_B T}{3h}$, where $F_s=0.1$ is a
reduction factor due to phonon scattering with defects in
nanoscale GNRs.$^{45,53)}$ Obviously, the optimization of $ZT$ at
finite $\kappa_{ph}$ situation is different from the case with
$\kappa_{ph}=0$ (see Fig. 7).

\begin{figure}[h]
\centering
\includegraphics[angle=0,scale=0.3]{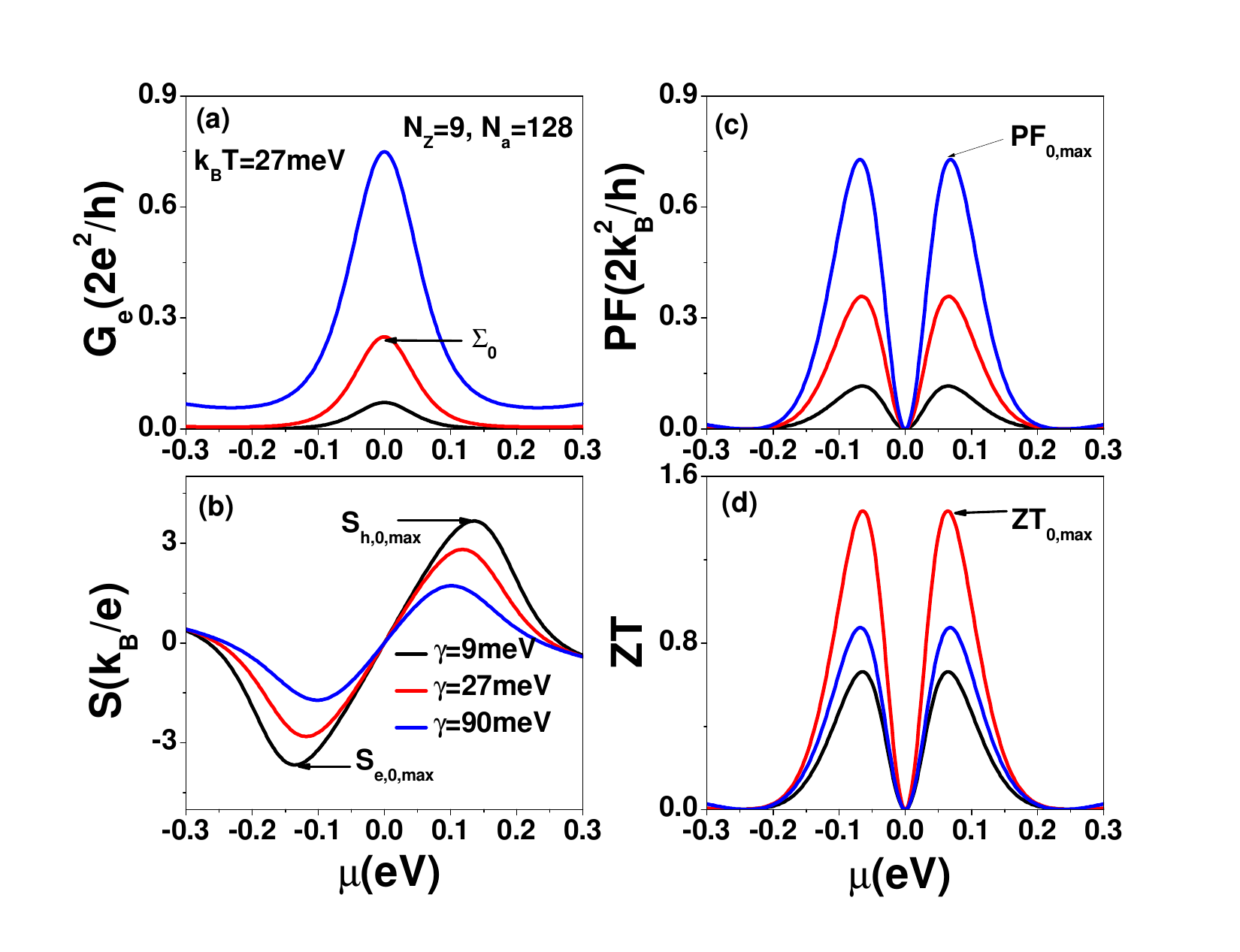}
\caption{(a)Electrical conductance, (b)Seebeck coefficient, (c)
power factor and (d) figure of merit as functions of chemical
potential for different tunneling rate ($\gamma_{L(R)}=\gamma$)
values at $k_BT=27meV$, $N_z=9$ and $N_a=128$. Note that
$\kappa_{ph}=F_s\frac{\pi^2k^2_BT}{3h}$ and $F_s=0.1$. }
\end{figure}

\section{Conclusion}
We have theoretically investigated the transport and
thermoelectric properties of GNRs with zigzag and armchair edges
in the framework of  Green's function approach within a
tight-binding model. GNRs can have sophisticate metallic or
semiconducting phases depending on the nanoribbon width. We
clarified quantum confinement effect on the transport and
thermoelectric properties of GNRs. In particular, we found that
the confinement effect of ZGNRs can highly enhance the Seebeck
coefficient at low temperature. When the zigzag sides are coupled
to the electrodes, electron transport through localize zigzag
states can be resolved only when the armchair length is smaller
than the decay length of zigzag edge states along armchair
direction. For GNRs in semiconducting phase in which the armchair
edges are coupled to the electrodes we can get significant
electron conductance through the zero-energy modes. For cases with
$N_a \gg N_z$ in Fig. 1(b), the top and bottom zigzag edges are
essentially decoupled and the thermoelectric behavior of
zero-energy modes can be well described by using two parallel
quantum dots with the same orbital degeneracy. Therefore, the
optimized $PF$ and the best $ZT_e$ of zero energy modes could be
analytically obtained in the case of $\kappa_{ph}=0$. We found
that $\Sigma_0$ of zero-energy modes is very robust against the
carrier scattering from point defects as long as they do not
appear on the edge. This mechanism provides the promising means to
reduce $\kappa_{ph}$ and remains the power factor of $PF_{0,max}$
resulting from zigzag edge states.

%\begin{flushleft}

%\end{flushleft}

%\begin{flushleft}
{\bf Acknowledgments}\\
{This work was supported by the Ministry of Science and Technology
(MOST), Taiwan under Contract No. 110-2119-M-008-006-MBK. The
author thanks Yia-Chung Chang for help with the manuscript
preparation and supporting IBM computer clusters.}
%\end{flushleft}
% 109-2112-M-001-046  and 110-2112-M-001-042
%\mbox{}\\

E-mail address: mtkuo@ee.ncu.edu.tw\\
%E-mail address: yiachang@gate.sinica.edu.tw\\

%\renewcommand{\thesection}{\mbox{Appendix s}} %\section{Appendix}~\Roman{section}
\setcounter{section}{0}

\renewcommand{\theequation}{\mbox{A.\arabic{equation}}} %\section{Appendix}
\setcounter{equation}{0} % reset counter

%\section{}
%\subsection{Derivation of the tunneling current formula using Dyson's equations\label{App:TC_l} }
\mbox{}\\

\appendix
\numberwithin{figure}{section}
\section{Electronic band structures}
Although the electronic band structures of GNRs have been
intensively studied,$^{30-36)}$ we briefly illustrate the band
structure of GNRs here to make this manuscript more readable. Fig.
2 shows the electron transport behavior from the left electrodes
to the right electrode via an GNR. Decoupling the electrodes, we
calculate the electronic band structures of AGNRs for different
$N_z$ values in Fig. A. 1, which exhibit the semiconducting and
metallic phases. $N_z=20$ shows a metallic phase. $N_z=10, 24, 30$
show semiconducting phases. Due to time-reversal symmetry,
electron-hole symmetry exists in Fig. A. 1. The results of Fig. A.
1 could explain the material phases shown in Fig. 2. {For the
cases of odd $N_z$ such as $N_z=5,7,9$ (not shown here), the
flat-bands appear at $E=\pm2.7eV$, which agree with analytical
solution of AGNRs.$^{33)}$

\begin{figure}[h]
\centering
\includegraphics[angle=0,scale=0.3]{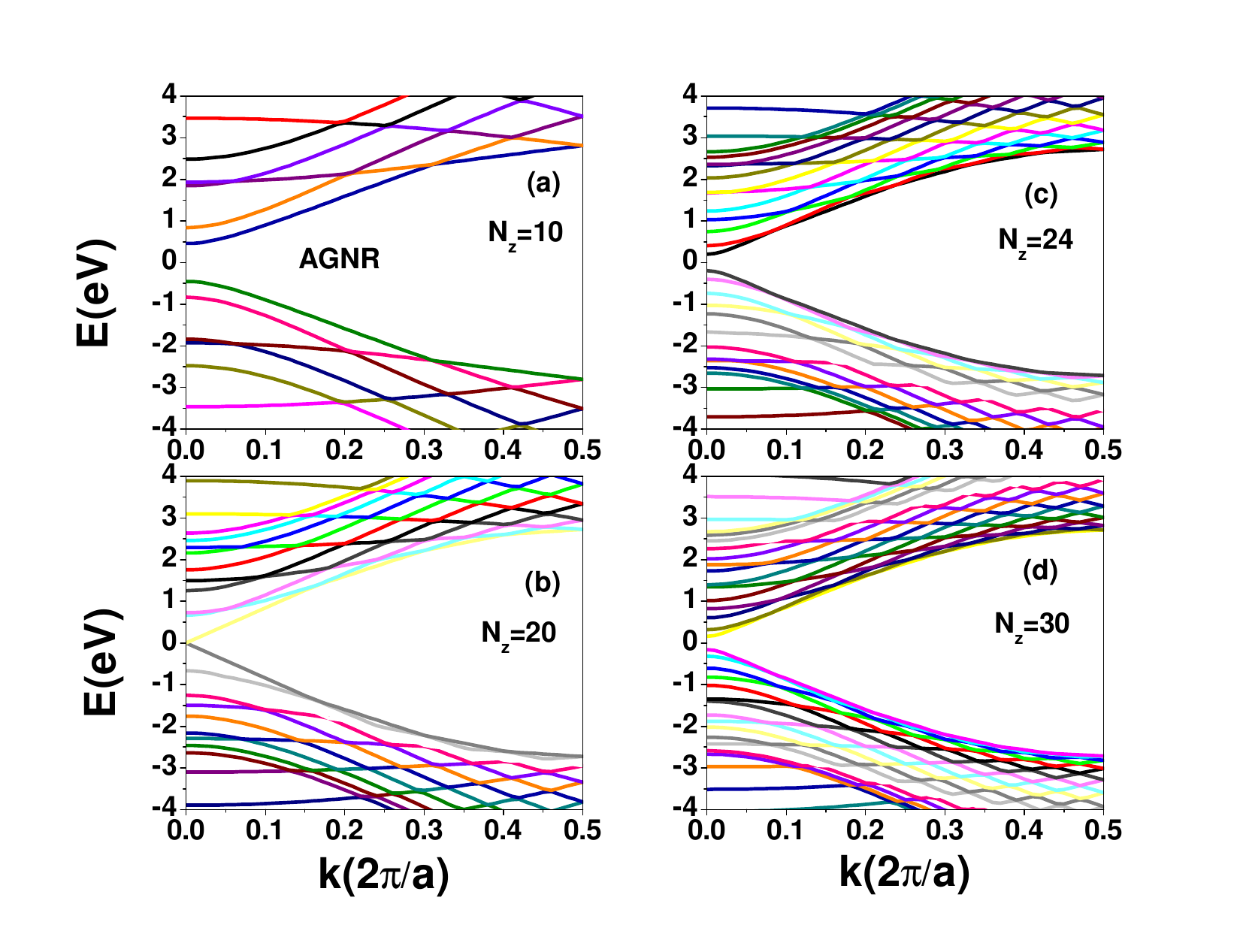}
\caption{Electronic band structures for armchair GNRs for
different $N_z$ values.}
\end{figure}

The extra peak $\Sigma_0$ shown in Figures 3 and 4 results from
the zigzag edge states. To demonstrate such a feature, we show the
electronic band structures of zigzag GNRs for different $N_a$
values. As seen in Fig. A. 2(d) with $N_a = 30$, there are a pair
of orbit degeneracy in the flat-bands with zero electron group
velocity. This zero energy flat-band modes is from
$k=\frac{2\pi}{3a}$ to $\frac{\pi}{a}$. In Ref.~[31] authors gave
an analytical solution for the large $N_a$ limit. The results of
Fig. A. 2(d) indicate that the zigzag edges are decoupled when
$N_a\ge 30$. For smaller $N_a$ values, only $k = \frac{\pi}{a}$
has zero energy modes. When $k$ is deviating from $\frac{\pi}{a}$,
the zigzag edges states form the bonding and antibonding states.
Therefore, zero-energy modes are lifted.$^{31)}$ Based on the
results of Figs. A. 1 and A. 2, we consider finite GNRs with
zigzag and armchair edges, which allow GNRs with zero-energy modes
protected by the band gap.$^{46)}$

\begin{figure}[h]
\centering
\includegraphics[angle=0,scale=0.3]{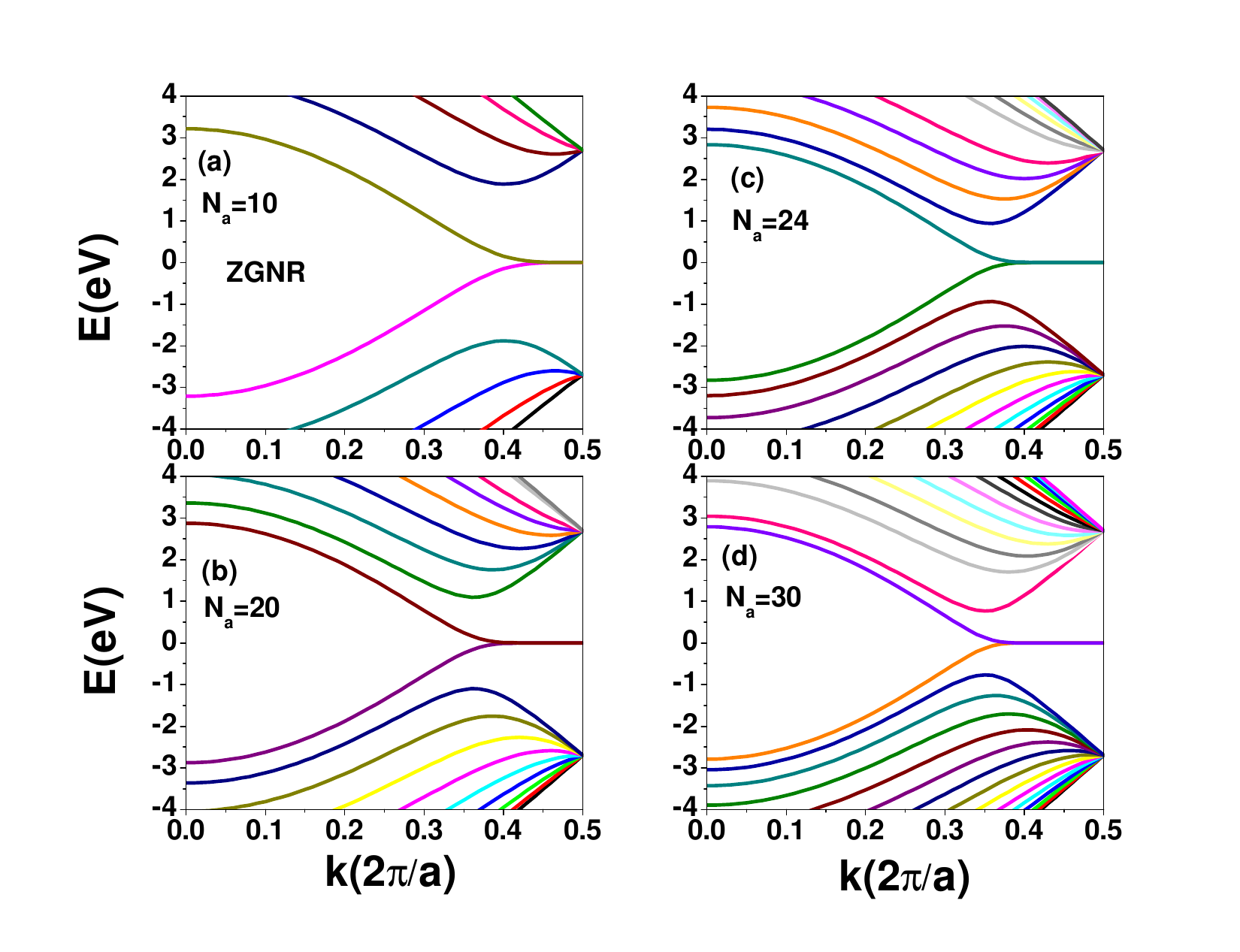}
\caption{Electronic band structures for zigzag GNRs for different
$N_a$ values.}
\end{figure}

\section{Charge density}
\numberwithin{figure}{section} To further reveal the properties of
edge states, we calculate the charge density of GNRs, which are
determined by the wave functions of finite size GNRs
($|\psi_{\ell,j}(\varepsilon)|^2$). The charge density of GNRs
with $N_a=100$ and $N_z=15$ decoupled with the electrodes is
plotted in Fig. B.~1 at various locations inside the GNR with
$(\ell,j)$ (see Fig. 1(a)). The maximum zigzag edge charge density
$|\psi_{\ell,j}(\varepsilon=0)|^2$ at $j=1$ decays quickly with
increasing lattice index for $j$ (j=3,5,7,9,11,13..). The results
show clear evidence for zero-energy modes resulting from the
zigzag-edge states with localized wave functions along armchair
edge direction. Such results of Fig. B. 1 explain why we can not
observe the zero energy modes in Fig. 2, but observe it in Fig. 3.

\begin{figure}[h]
\centering
\includegraphics[angle=0,scale=0.3]{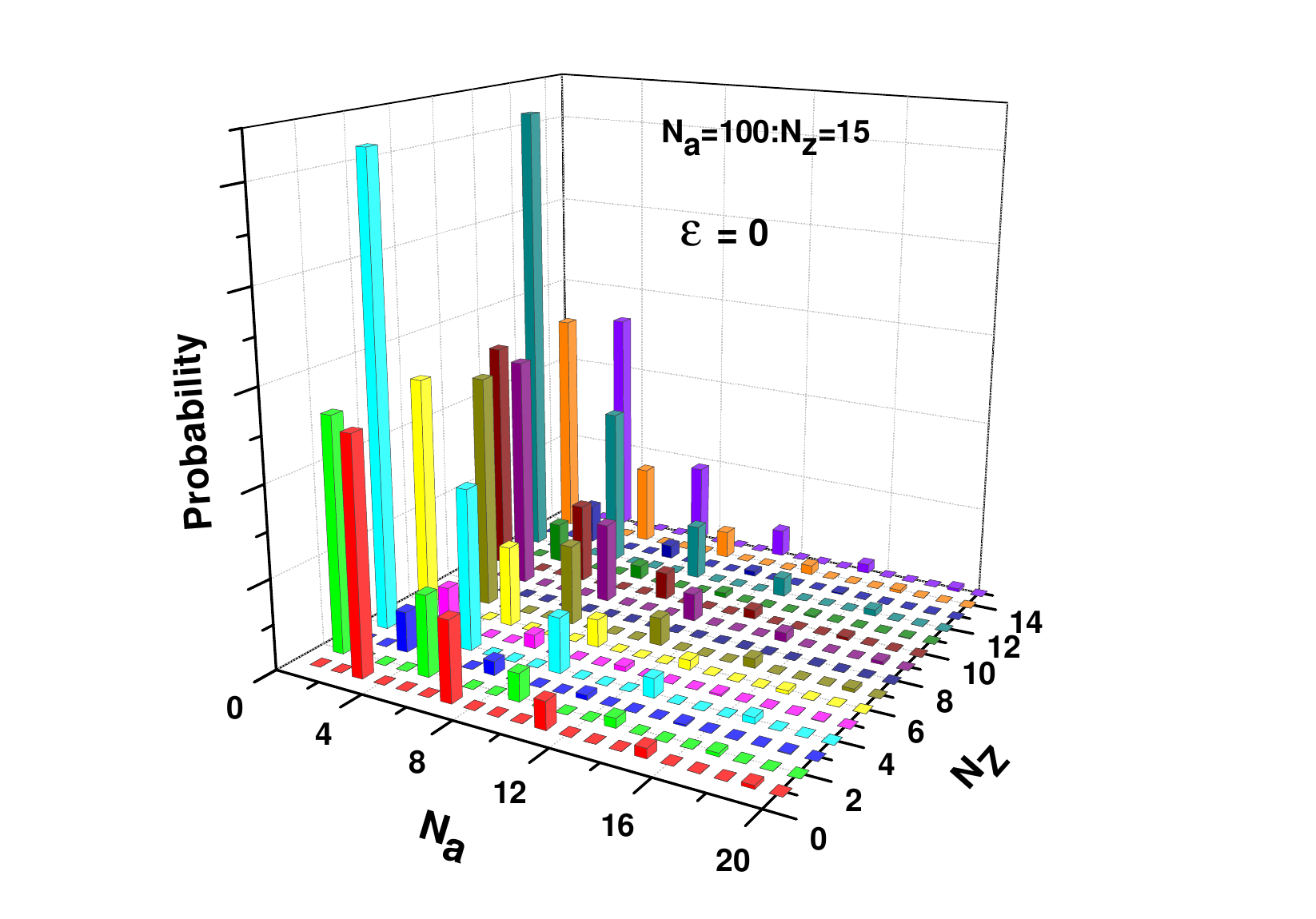}
\caption{Charge density of GNRs with $N_a=100$ and $N_z=15$. We
plot $|\psi_{\ell,j}(\varepsilon=0)|^2$ for $\ell$ from 1 to 15
and $j$ from 1 to 20.}
\end{figure}

Next, the charge density of GNRs with $N_z=121$ and $N_a=24$ are
plotted in Fig. B.~2 at various locations of the GNRs. The maximum
zigzag edge charge density occurs at $j=1$ and $j=24$ for even
$\ell$ numbers. Due to a small $N_a$ value, the localized edge
states ($\varepsilon=0.068eV$) form the bond and antibonding
states. As seen in Fig. B.~(2a), the oscillatory charge density
along the direction of zigzag edges. For the case of
$\varepsilon=0.58eV$, the charge density of GNRs shows delocalized
wave functions.

\begin{figure}[h]
\centering
\includegraphics[angle=0,scale=0.3]{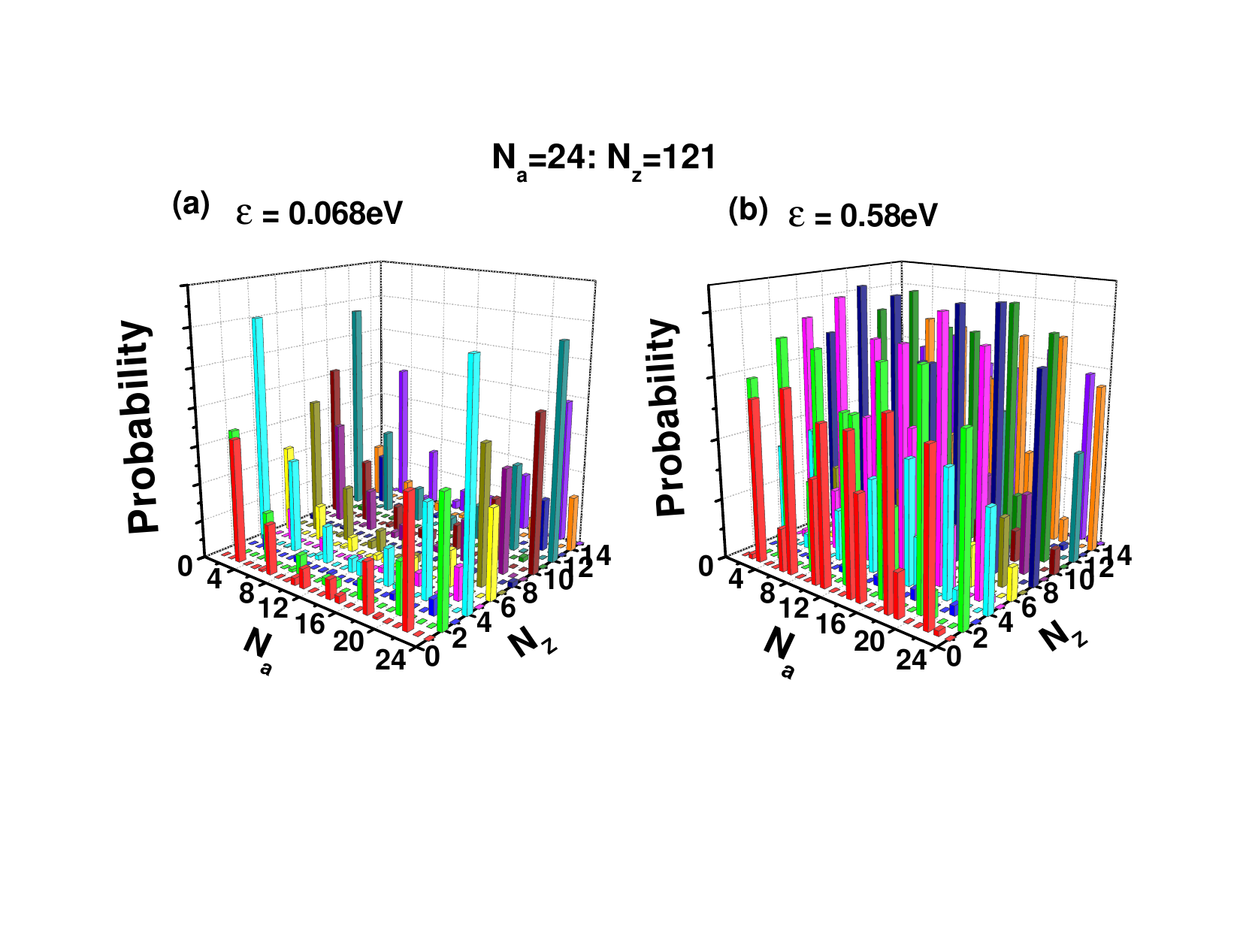}
\caption{Charge density of GNRs with $N_z=121$ and $N_a=24$. (a)
$|\psi_{\ell,j}(\varepsilon=0.068eV)|^2$, and (b)
$|\psi_{\ell,j}(\varepsilon=0.58eV)|^2$. $\ell$ is accounted from
1 to 14 and $j$ is from 1 to 24.}
\end{figure}

\end{document}